# A Thermodynamic Core using Voltage-Controlled Spin-Orbit-Torque Magnetic Tunnel Junctions

Albert Lee[1], Bingqian Dai[1], Di Wu[1], Hao Wu[1], Robert N. Schwartz[1], and Kang L. Wang[1,2,3],

[1]Department of Electrical and Computer Engineering, UCLA, Los Angeles, CA, 90095, United States of America

[2]Department of Physics and Astronomy, UCLA, Los Angeles, CA, 90095, United States of America

[3]Department of Material Science and Engineering, UCLA, Los Angeles, CA, 90095, United States of America

E-mail: oncefriends9206@gmail.com, klwang@g.ucla.edu

## Abstract

We present a magnetic implementation of a thermodynamic computing fabric. Magnetic devices within computing cores harness thermodynamics through its voltage-controlled thermal stability; while the evolution of network states is guided by the spin-orbit-torque effect. We theoretically derive the dynamics of the cores and show that the computing fabric can successfully compute ground states of a Boltzmann Machine. Subsequently, we demonstrate the physical realization of these devices based on a CoFeB-MgO magnetic tunnel junction structure. The results of this work pave the path towards the realization of highly efficient, high-performance thermodynamic computing hardware. Finally, this paper will also give a perspective of computing beyond thermodynamic computing.

## 1. Introduction

The relevance of thermodynamics in modern computer science/technology has its roots in the seminal work of Landauer [1] who provided insight into the thermodynamic limits on erasure of information for a computer system operating in an irreversible manner; put more succinctly, Landauer concluded that logically irreversible information processing incurs an unavoidable thermodynamic cost. Landauer, in collaboration with Bennett [2], developed the fundamental basis of the thermodynamics of reversible computing. The basic idea behind reversible computing is that information within the computing machine is transformed reversibly in such a manner that the energy associated with the information can be recovered and reused in subsequent computational steps without being dissipated as heat. The more recent focus on recasting computing in terms of thermodynamics is related to the challenges of power consumption, device fluctuations, fabrication costs, and computer architecture; all of which, at the basic level, are fundamentally thermodynamic in nature. This paper will also give a perspective of computing beyond thermodynamic computing.

Further improvements in computer technology will require extensive modifications in many aspects; however, we focus here on the issue of energy consumption. In other words, to make computers more functionally efficient, then we need to focus on the energy required to effectively induce a state change. As a guide to minimizing energy consumption and dissipation in potential devices that could provide a platform for future computers, researchers have long considered that Nature's computational capacity may provide a pathway to achieve this goal. At the most fundamental level, addressing computing in terms of Natural dynamics is not new; specifically, for many decades living systems have been under scrutiny in order to reveal the underlying principles responsible for their energy-efficient and computational capabilities, which far surpass our current silicon-based technologies. The discipline called Thermodynamic Computing, for example, addresses this important issue. It is noted as a type of computing that sits between Classical Computing and Quantum Computing in reducing energy dissipation per compute cycle and moves to increasing the number of computations per device count.

Compared to software approaches, utilizing intrinsic physical properties of devices for information processing greatly speeds up the computational process. Within the framework of thermodynamic computing, specifically for our interest here, naturally occurring device-level fluctuations will be exploited so as to populate the state space of a much smaller number of devices compared with that needed for today's von Neumann computers, and thus with low-energy representations that will yield efficient state changes with accompanying low energy consumption and low heat dissipation. It is anticipated that these devices will not only provide a direct mapping to algorithms sharing the same operation rules, but also capture secondary effects such as various thermal noise sources not included in simple models. For example, Boltzmann Machines [3], a general computing medium that finds use cases in applications ranging from pattern recognition, combinatorial optimization, and neural behavior modelling [4] to NP-complete problems [5], exhibit thermodynamical properties in both training and search modes [6]. Boltzmann machines also benefit from thermal annealing algorithms [7], in which temperature is introduced into the system via the parameter $T$, which scales down weights and energies. Through a decaying temperature schedule, the system speeds up the search towards the minimum energy state by going from a global random search to one of local optimization [7].

In this work, we exploit the use of spin phenomena to model thermodynamic evolution. Our Magnetic Thermodynamic Core (MTC) provides stochastic dynamics governed by the temperature and biased through a learning process. Subsequently, we show the application of our MTC system in simulating a Boltzmann machine. The MTC is essential for the development of new cognitive computing paradigms based on thermodynamic evolution. The remainder of the paper is organized as the follows: Section 2 describes the MTC, the employed magnetic device, and its dynamics. Section 3 presents simulations of the device and verifies its dynamics. Also presented in this section is the simulation of a 4-neuron Boltzmann machine on the MTC system. Section 4 presents experimental results on the realization of the device, and Section 5 concludes this paper with a perspective.

## 2. Magnetic Thermodynamic Core

A thermodynamic computing architecture [8,9] is composed of:

(1) An interconnected fabric of cores, each containing stochastic states and/or functions;
(2) The connections and states may be constrained by design or dynamically learned;
(3) The network evolves from a dynamic state to an equilibrium state with low energy as the solution.

The discussion of the above three elements follow. We propose a magnetic implementation of such an architecture, using the thermal fluctuations of a magnetic tunnel junction (MTJ) in combination with two spin-control knobs: the Voltage-Controlled (VC) Magnetic Anisotropy (VCMA) effect [10,11], which acts as a temperature parameter controlling the thermal stability; and the Spin-Orbit Torque (SOT) effect [12,13], which allows network configurations to affect the direction of evolution. The device is employed with peripheral circuitry to create the MTDC. Figure 1 shows the basic structure of the device, which consist of a VC-MTJ resting on top of a heavy metal bus. The VC-MTJ has two magnetic layers separated by a tunneling barrier. The top magnetic layer has a relatively high coercivity and fixed magnetization during operation (fixed layer), while the bottom magnetic layer can freely change its magnetization under different electrical and magnetic conditions (free layer). The SOT bus is composed of materials that scatter spins to different locations along the bus, such as heavy metals and topological insulators. In this, a spin current that flows to the top of the bus will couple to the free layer's magnetization.

MTJ devices have the unique ability to harness thermal fluctuations, an operation which is extremely costly for CMOS or other memristive devices [14,15]. The combination of VC and SOT phenomena in an MTJ is critical for modelling thermodynamic evolution: A VC-MTJ can model temperature, yet it does not have the ability to configure low-energy system states. SOT-only devices and STT-based designs such as p-bits [16] possess the ability to tune state distribution, but are unable to model temperature. On the other hand, spin-oscillator and spin-wave computing methods use coupling and frequency synchronization/locking for optimization and generally attempt to ignore thermal influence [17,18].

In our perpendicularly magnetized device, the fixed layer magnetization is pointing in the $+z$ direction, whereas the free layer can point in either the $+z$ or $-z$ direction. When the magnetization of the two layers are aligned ($+z$ and $+z$), the MTJ is said to be in a parallel (P) state, displaying a low resistance; while an anti-aligned magnetization (i.e. $+z$ and $-z$) is in the anti-parallel (AP) state with high resistance. When the current through the heavy metal bus is in the $+x$ direction, a $+y$ spin current moves to the top of the bus; while a current in the $-x$ direction causes a $-y$ spin current to move upwards. In Sections 2.1 through 2.4 below, we will first describe each of the magnetic phenomena that enables the MTC; the core itself will be introduced in Section 2.5

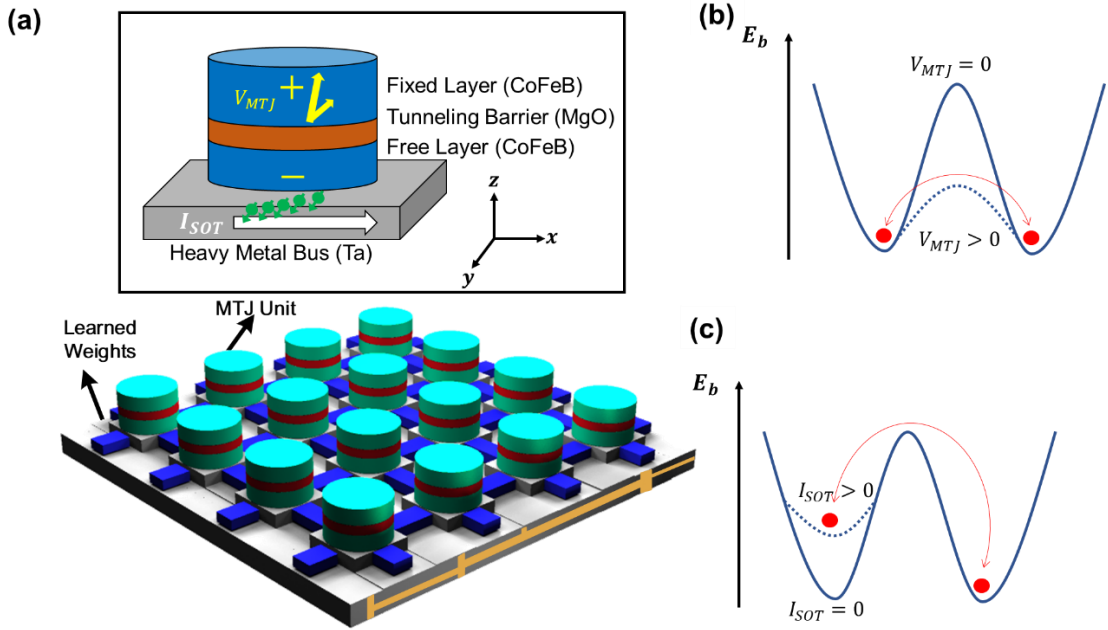

**Figure 1.** (a) Illustration of the MTC system, utilizing an array of magnetic devices to generate thermodynamics. Inset: the magnetic device for each core, composed of a VC-MTJ in direct contact with a SOT bus. (b) A double potential well showing two energy minimum states of the MTJ. Thermal fluctuations cause random flipping in the device state with an average frequency dependent on the energy barrier. The energy barrier can be modulated through the VCMA effect by an applied voltage. (c) The current flowing through the device's SOT bus, which is determined by the connections between cores, biases the potential of each state.

### 2.1. Thermal fluctuation

The state of an MTJ is susceptible to random switching arising from thermal fluctuations. The frequency in which the device state fluctuates is dependent of the energy barrier $E_b$ between the two states of the MTJ, and is characterized by the *thermal stability* $\Delta = \frac{E_b}{k_b T}$, where $k_b$ is the Boltzmann constant and $T$ is the temperature. The thermal stability gives a measure of the time constant $\tau$ in which the device retains its state (retention time) under an external bias field through the relationship [19] :

$$\tau = \tau_0 \, exp\left(\Delta - \frac{M_s H_B t_{fl} A}{k_b T}\right), \quad (1)$$

Here, $M_s$ is the saturation magnetization, $H_B$ is the magnitude of the bias magnetic field, $t_{fl}$ is the thickness of the free layer of the tunneling barrier, and $A$ is the area of the MTJ. In the case of a free layer with perpendicular anisotropy, the anisotropy is due to the spin orbit coupling (SOC) at the interface. The effective thermal stability is determined by the strength of its perpendicular magnetization anisotropy (PMA), under the influence of an external field created by its demagnetization in Equation 1 [19,20]:

$$\Delta' = \frac{(K_i - 2\pi(N_z - N_y)M_s^2 t_{fl})A}{k_b T}, \quad (2)$$

where $K_i$ is the PMA coefficient and $N_{x,y,z}$ is the demagnetization factor along each axis. The fluctuations in an MTJ directly capture the thermal dynamics in a realistic physical system.

## 2.2. Voltage-controlled Magnetic Anisotropy

In specially designed MTJs, an applied voltage can alter the magnetization of its magnetic layers. This effect, called the voltage-controlled magnetic anisotropy effect, can be induced by purely electronic effects [21,22], chemical effects [23,24], or mechanical effects [25,26] These effects originate from the hybridization of atomic orbitals, and cause a modification of $K_i$ via the change in strength of the SOC by the voltage across the MTJ at the free layer-tunneling barrier interface [11,27]. This can be put in the form of [19]:

$$K_i' = K_i - \frac{\xi V}{t_{fl}}, \quad (3)$$

where $\xi$ is the VCMA coefficient and $V$ is the applied voltage. Through the VCMA effect, an applied voltage linearly modulates the thermal stability of the device. Substituting Equation 3 into Equation 2, the thermal stability can be explained as:

$$\Delta(V) = \frac{\left(K_i - \frac{\xi V}{t_{fl}} - 2\pi(N_z - N_y)M_s^2 t_{fl}\right)A}{k_b T}. \quad (4)$$

A higher applied voltage reduces the device's thermal stability, creating the same effect as increasing the temperature of the device in equation 2.

## 2.3. Spin-orbit torque effect

Due to the spin Hall effect (SHE), electrons flowing in the $+x$ direction through the heavy metal bus generate a proportional spin current in the $+z$ direction with a spin polarization in the $+y$ direction. This spin current gives rise to a spin accumulation at the interface between the free magnetic layer and the heavy metal bus. This spin current gives rise to a spin accumulation at the interface between the free magnetic layer and the heavy metal bus. The spin accumulation can induce a torque that acts on the adjacent free magnetic layer and gives rise to the dynamics of the magnetization via the Landau–Lifshitz–Gilbert (LLG) equation. This creates an effective torque [28] of:

$$\vec{\tau}_{SOT} = \gamma \frac{\hbar \theta_{SOT} J}{2e\mu_0 M_s t_{fl}} \, (\hat{m} \times (\hat{\sigma} \times \hat{m})). \quad (5)$$

Here, $\hat{m}$ is the unit vector along the direction of magnetization, $\hat{\sigma}$ is a dimensionless vector along the spin polarization direction, $\gamma$ is the electron gyromagnetic ratio, $\hbar$ is the reduced Planck constant, $\theta_{SOT}$ is the spin orbit torque angle that describes the conversion efficiency between electrical and spin current, $J$ is the electrical current density, $e$ is the unit electron charge, $\mu_0$ is the permeability in vacuum, $M_s$ is the saturation magnetization, and $t_{fl}$ is again the thickness of free layer. This torque biases the MTJ state proportional to the amount of SOT current ($I_{SOT}$), allowing configurable connections (weights) to affect the evolution of the MTC during its operation.

*2.4. Voltage Dynamics*

Electrically, an MTJ acts as a capacitor (C) in parallel with a resistor (R) [29], in which the capacitance originates from the insulating barrier between the two magnetic layers and the resistance from the current through the tunneling barrier. The capacitance scales inversely proportional to the thickness of the barrier, whereas the resistance exhibits an exponential dependence on the barrier thickness because of tunneling. A simple expression for the dynamic evolution of an initial voltage placed across the MTJ follows a RC decay:

$$\frac{\partial V_{TJ}(t)}{\partial t} = \frac{V_{TJ}(t) - V_o}{RC}, \tag{6}$$

where $V_0$ is the equilibrium voltage and $V_{TJ}$ is the voltage across the MTJ. The RC model is well established to describe the first-order thermal dynamics, in which the temperature of a body evolves towards its ambient temperature.

*2.5. Magnetic Thermodynamic Core*

The MTC, shown in Figure 2, utilizes the VC-SOT device to create a thermodynamic evolution and generate a distribution of states based on the arriving current of SOT. The MTC takes in the states of other connected MTCs as inputs, where a digital weighted sum module computes a weighted sum of the inputs and drives the $I_{SOT}$ through the bus of the VC-SOT device. A thermal module supplies the initial value of $V_{TJ}$ across the tunneling barrier of the MTJ, which then discharges through each MTJ device to model the first-order temperature evolution. A custom temperature schedule can also be adopted. The state detector detects the state of the MTJ, which is then sent to inputs of other MTCs.

The average switching frequency ($f_{sw}$), obtained by substituting Equations (2) and (3) into (1), can be represented as

$$f_{sw} = \tau^{-1} = \alpha_1 e^{-\alpha_2 V_{TJ}(t)}, \tag{7}$$

where $\alpha_1$ and $\alpha_2$ are collections of constants. The time evolution of $V_{TJ}$, following Equation (6), can be written as

$$\frac{\partial V_{TJ}(t)}{\partial t} = \beta\big(V_{TJ}(t) - V_o\big), \tag{8}$$

where $\beta$ is also composed of constants. As stated in section 2.3, $V_{TJ}$ is an indicator of the temperature of the core. The distribution of P and AP states is proportional to the switching probability (or write error rate) of the MTJ. Empirically, the switching probability is exponentially dependent of the write current once it exceeds the critical current. Thus, the distribution of AP and P states can be expressed as

$$\frac{P_{AP}}{P_P} = \frac{e^{\gamma_1 + \gamma_2 I}}{1 - e^{\gamma_1 + \gamma_2 I}}, \tag{9}$$

where $I$ is the current driven through the heavy metal, and $\gamma_1, \gamma_2$ are constants.

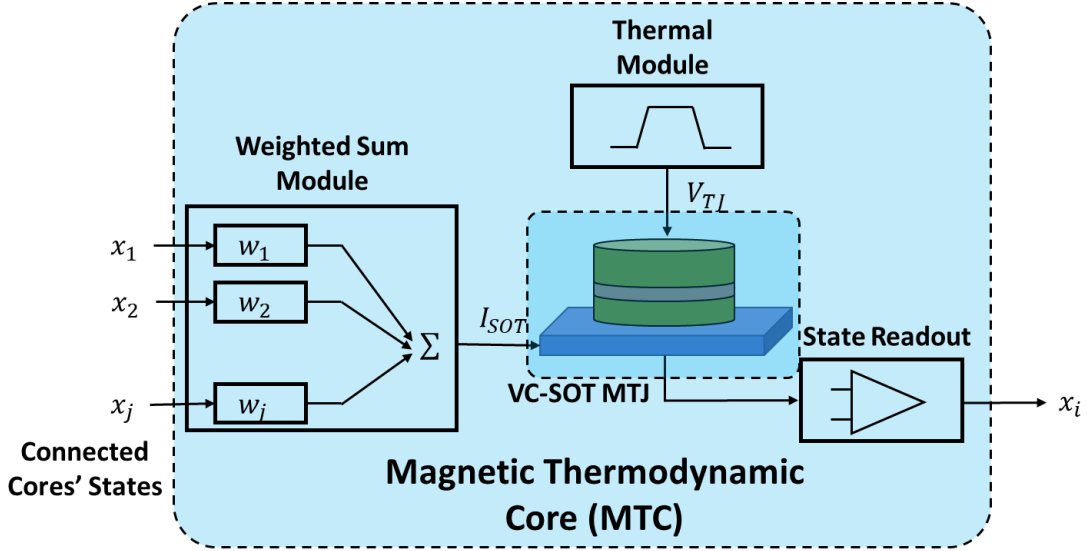

**Figure 2**. Schematic of the MTC. The MTC takes in the states of other connected modules as inputs. The weighted sum module computes a weighted sum $\sum x_i w_i$ of the MTC's inputs, and drives a current through the SOT bus accordingly. The thermal module places an initial voltage across the MTJ, which evolves following the first-order thermodynamics. The state detector detects the state of the MTJ and sends it to other connected cores. Weights may be programmed or learned throughout the computing process.

## 3. Simulation of the MTC system

In this section, we first verify the MTC dynamics in Section 2 through simulation, and then show the application of the MTC system for computing the minimum energy states of a Boltzmann Machine. For the MTJ model, we adopt the LLG based macro-spin model with thermal noise [19] as the backbone. This model describes the dynamics of the free layer's magnetization under an effective field, generated by the combination of the PMA, VCMA, demagnetization field, and external field. We combine this with the SOT model described in [28], which models the effect as an additional current-dependent damping torque (Equation 5). The VCMA coefficient 80 fJ $V^{-1}$ $m^{-1}$ and spin hall angle of 1 are used, and other parameters are adopted from our experiments in [19,30].

The simulated result of the relationship between the retention time and MTJ voltage is shown in Figure 3. Due to the exponential relationship between voltage and retention time, we display the temporal range of highest computing interest (nanoseconds to ~microseconds). The solid line with dots on the log-linear plot follows the exponential relationship of Equation 7. A voltage of 0.5 V corresponds to an average retention time of ~20 μs, and at 0.6 V it reduces to less than 100 ns. The SOT-induced state bias is shown in Figure 4. In this simulation, the MTJ is initialized to a random state. Afterwards, a voltage pulse of 0.5 V, 50 ns is applied on the MTJ, and a current pulse flows through the SOT bus. At the end of the pulse, the amount of P and AP states are recorded. The curve follows the relation in Equation. 9.

For simulation of the Boltzmann Machine, we compose the network of stochastic binary neurons ($S_i \in \{0,1\}$) with all-to-all, symmetric ($W_{ij} = W_{ji}$) connections. Each neuron computes its potential $Z_i$ as a weighted sum of all other neuron states ($Z_i = \sum_{j \neq i} W_{ij} S_j$), and stochastically updates its state with probability $P(S_i = 1) = (1 + e^{-Z_i})^{-1}$. The system aims to minimize its Hamiltonian, or energy, given by $E = -\sum_{i<j} W_{ij} S_i S_j$. The implementation of our magnetic thermodynamic computing system is shown in Figure 4(a). Each neuron in the Boltzmann machine maps to a single MTC in our computing system. For neuron $i$, its state $S_i$ is represented by the MTJ's state ($AP\ state: S = +1;\ P\ state: S = 0$), and its connected weights $W_{ij}$ are stored in the weighted sum module. Its inputs $x_i$ are the states of other connected neurons $S_j$. The weighted sum module drives a current through the SOT bus via the relationship $I_{SOT,i} = i_0 \sum_{j \neq i} W_{ij} S_j$, where $i_0$ is a device-design dependent current scalar.

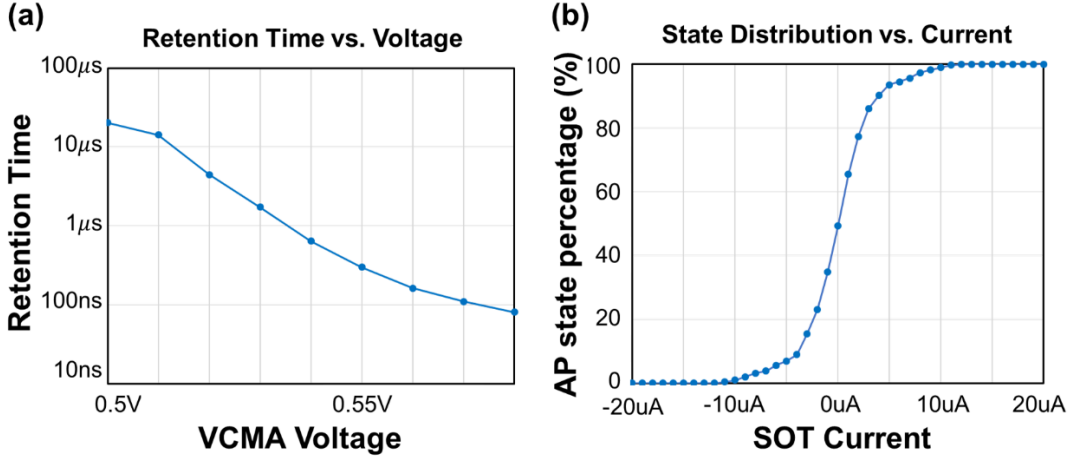

**Figure 3**. Dynamics of the VC-SOT device (a) Voltage-retention time relationship calculated from the LLG equation, showing a linear relationship in the semi-log plot; and (b) SOT induced probability bias. These relationships follow those derived in Section 2.

The result of the complete energy landscape of a 4-neuron Boltzmann Machine is shown in Figure 4(b). Here, we consider weights of $W_{ij} = \pm 1$ to represent both positive and negative connections. The *x*-axis (spin configuration) and *y*-axis (weight configuration) of the landscape represent binary-coded values: for example, a spin configuration of 9 represents the neuron's state configuration of $S_{i;\ i \in [1,4]} = \{1,0,0,1\}$; and a weight configuration of 16 means a weight combination of $W_{ij;\ i<j;\ i,j \in [1,4]} = \{-1,+1,-1,-1,-1,-1\}$. When the weight between two neurons is positive ($+1$), the lowest energy occurs when both neuron states are $+1$; while for a negative weight ($-1$), this state gives instead the highest energy. Note that for spin configurations $0 = \{0,0,0,0\}$, $1 = \{0,0,0,1\}$, $2 = \{0,0,1,0\}$, $4 = \{0,1,0,0\}$, and $8 = \{1,0,0,0\}$, the energy is always 0 regardless of the weight configuration, since $S_i S_j = 0$ for all combinations of $i, j$.

Figures 5(a) and 5(b) show the final system state distribution from 1,000 trials under two weight configurations: one with a single global minimum ($W_{ij} = \{+1, +1, +1, -1, +1, +1\}$), and one with five equal-potential minima ($W_{ij} = \{-1, -1, -1, -1, -1, -1\}$), along with their energy landscape and network structure. In all simulations, neuron states $S$ are initialized to $+1$ or 0 randomly, and the initial voltage across the MTJ is set to 0.6 V. Figures 5(c) and 5(d) show the trajectories of a random simulation trial; where we observe that initially, at high temperatures, the system states jump between several local minimum, then eventually settle to the global minimum. We successfully verify that the MTC system can find the minimum energy state with ~99% probability within 300 ns. The network can be scaled to larger numbers of neurons by increasing the capacity of the summation module. It can also be changed to structures such as the multi-layer or restricted Boltzmann machines [3], by connection design.

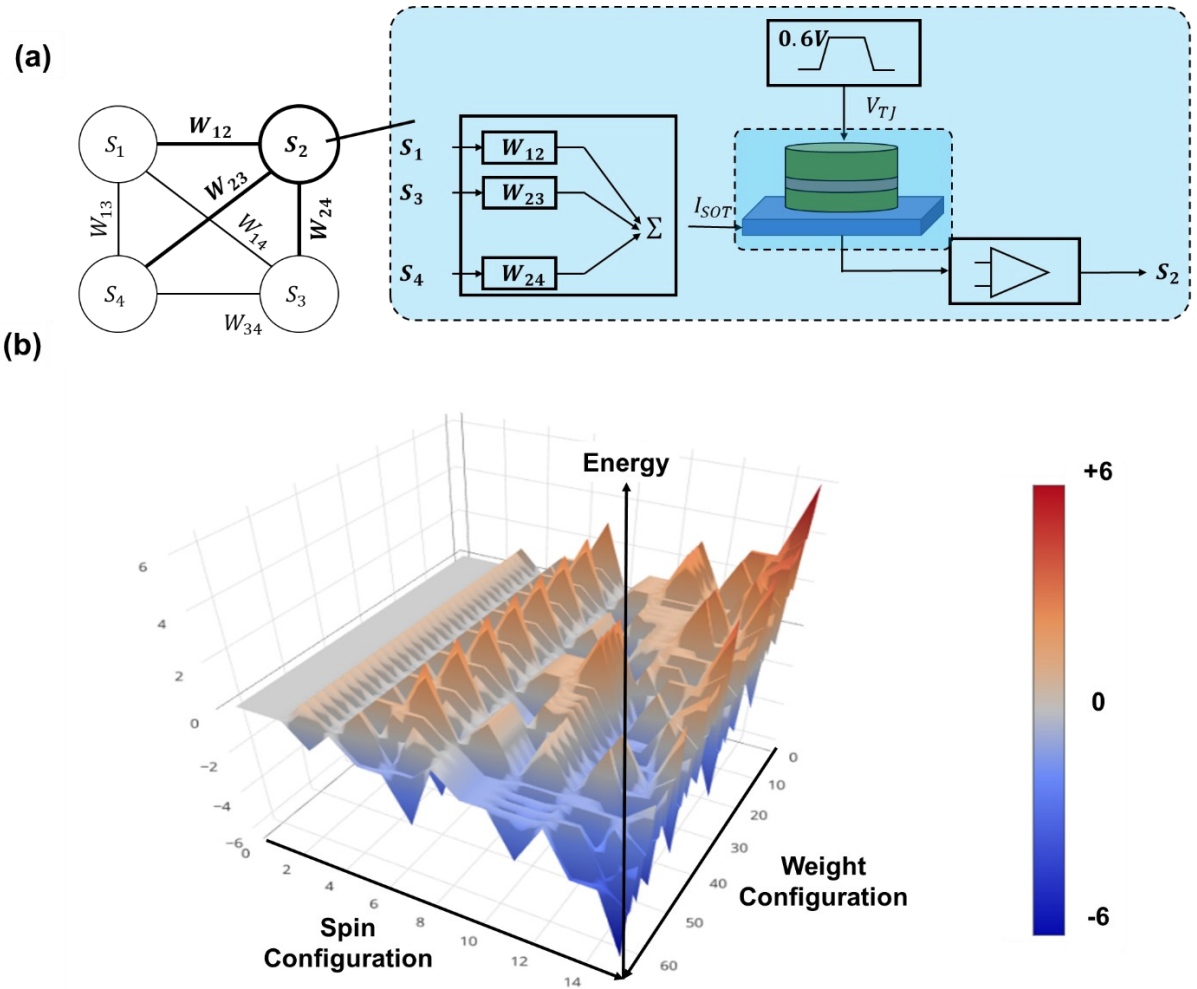

**Figure 4.** (a) Simulation setup of a 4-neuron Boltzmann machine on our MTC system. Each neuron is mapped to one MTC. For neuron $i$, its state $S_i$ is represented by the MTJ's state, and its connected weights $W_{ij}$ are stored in the weighted sum module. Its inputs $x_i$ are the states of other connected neurons $S_j$ (b) Energy landscape of the 4-neuron Boltzmann machine, with binary-coded spin configurations and weight configurations.

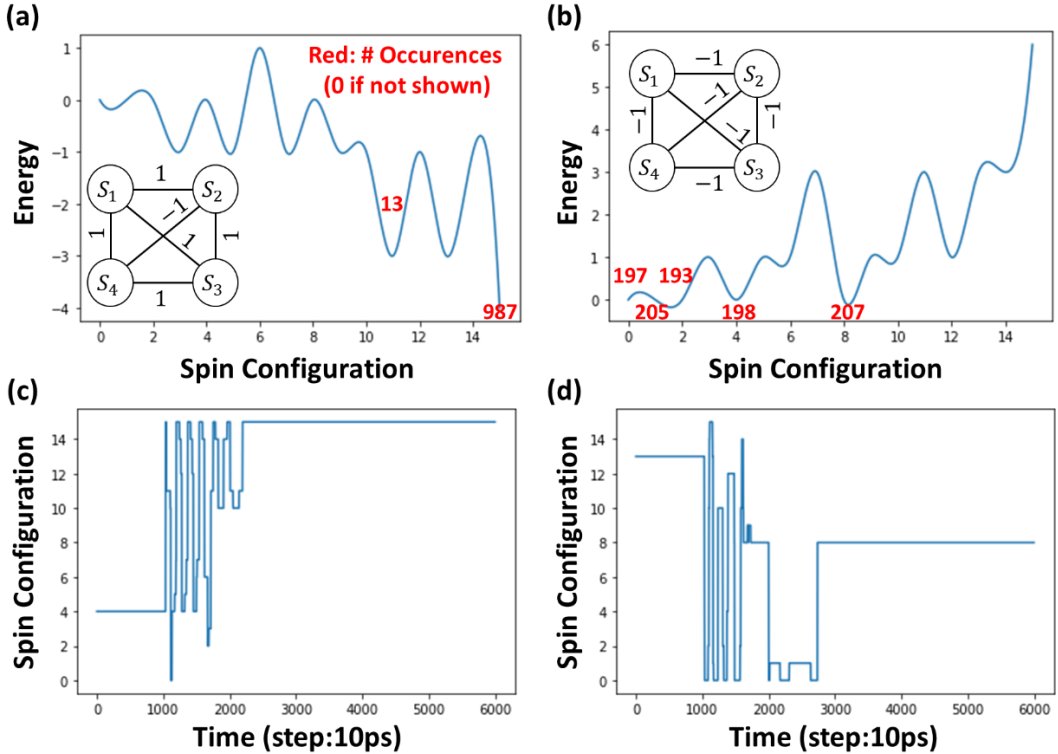

**Figure 5.** Simulation results of the 4-neuron Boltzmann machine on the MTC system. (a,b) Smoothed energy landscape and network structure of (a) a weight configuration with a single global minimum, and (b) a weight configuration with multiple local minima. Red numbers on top of the spin configuration axis indicate the number of occurrences of final states out of 1000 runs. (c, d) Trajectories of a random simulation for the two configurations. We observe a global search when the temperature is high, and settles toward the global minima as it decays.

## 4. Experiments

In this section, we experimentally demonstrate the device principles (SOT, VCMA, and thermal fluctuations) of the MTC in a CoFeB-MgO MTJ structure as neurons used in Section 3, where CoFeB is the material for the free and fixed layers and MgO the tunneling barrier. For sake of fabrication and measurement, we measure the SOT and VCMA effects through Hall bar structures and measure the thermal stability on MTJ devices. These experiments are critical to the realization of the TMC system.

### 4.1. SOT Switching

We demonstrate the SOT effect of Figure 2 with a tantalum heavy metal bus (Ta(5)/CoFeB(0.9)/MgO(2)) as shown in Figure 6 (a), in which the parenthesis after each material indicate the thickness of each layer in nanometers. The stack is deposited by magnetron sputtering on a thermally oxidized wafer at room temperature. The film stacks were then annealed at 200 °C for 30 min to enhance their PMA, and patterned into 20 x 130 μm Hall bar devices through photolithography, dry etching, and electron beam lithography techniques. The magnetic hysteresis loop of the device is shown in Figure 6 (b), obtained by measuring the anomalous Hall effect (AHE) voltage when a longitudinal current in the $x-$direction is applied to the current channel. We observe that the coercivity of the device is 100 Oe and the AHE resistance is 2 Ω.

Figure 6 (c) shows the SOT switching of the device. To couple the $+y$ spins generated by the SOT bus to the perpendicularly ($\pm z$) magnetized layers, an $x$−direction in-plane magnetic field of +20 Oe/-20 Oe is applied. Current pulses between 10 to ~20mA are able to switch the magnetization from down ($-z$) to up ($+z$) and vice versa. In a complete VC-SOT MTJ device, this in-plane external magnetic field could be generated by another adjacent in-plane ferromagnetic layer (such as a thick CoFeB (2-3 nm) layer) [31,32]. The additional in-plane ferromagnetic layer provides a built-in in-plane stray field that is localized at the free layer and thus, enables SOT switching without an externally applied field. There are also alternatives to achieve this, such as replacing the external bias field by the exchange bias from an anti-ferromagnetic under layer [33], a wedge-shaped ferromagnetic free layer that achieves field free SOT switching [34], or inter-layer exchange coupling via the RKKY interaction [35].

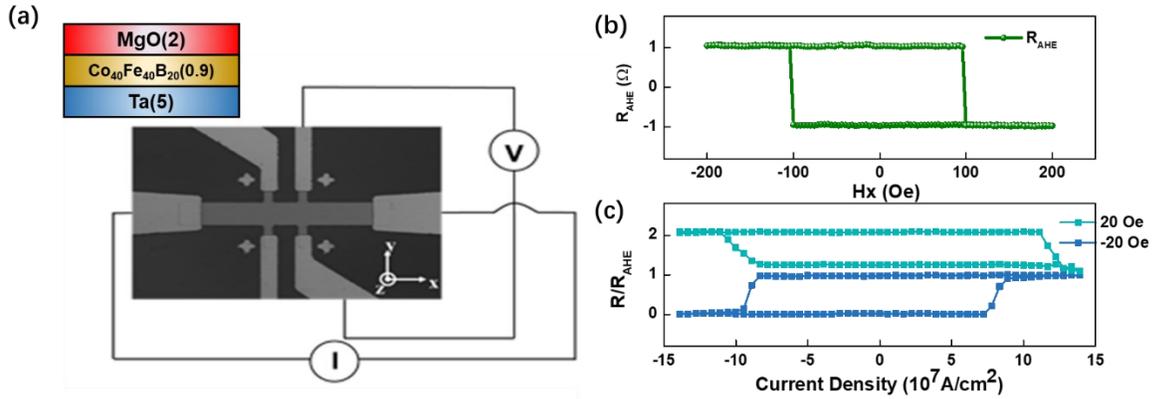

**Figure 6.** (a) Material stack of the Ta(5)/CoFeB(0.9)/MgO(2) Hall bar, and the microscope image of the 20 x 130 m Hall bar device. During the measurement, the applied current goes through the longitudinal channel of the Hall bar and the voltage is measured from the transverse channel. (b) The magnetic hysteresis loop and (c) SOT switching of the magnetization.

*4.2. VCMA and Thermal Stability*

We demonstrate enhanced VCMA effect using an Ir/Mo seed layer (Ir(5)/Mo(1)/CoFeB(0.9)/MgO(2)), as shown in Figure 7(a). Similarly, the stack is deposited by magnetron sputtering, annealed at 400℃ for 30 minutes, and fabricated into Hall bar structures. To extract the VCMA coefficient, we measured the magnetic hysteresis loop under different gate voltages (Figure 7(b)) and extract the PMA coefficient $K_i$ under different voltages (Figure 7(c)). The $K_i$-Electric Field plot reveals a VCMA coefficient $\xi$ of -51.8 fJ V$^{-1}$ m$^{-1}$.

To demonstrate the voltage-controlled thermal stability, we fabricate the VCMA stack into MTJs (Ir(5)/Mo(1)/CoFeB(1.1)/MgO/CoFeB(1.3)) as shown in Figure 8 (a). The MTJ is 80 nm in diameter with resistance-area product (RA) of ~200 Ω μm$^2$, and the hysteresis loop (Figure 8 (b)) displays a 180 % ratio between the P and AP states. The thermal stability Δ under different applied voltages is shown in Figure 8 (c), where we observe a linear change in Δ with the applied voltage matching the results reported in [25,26]. Based on this plot, we expect the PMA to decrease to near 0 with an applied voltage of 1.4 V. At this voltage, the low thermal stability (high MTC temperature) results in a device state purely resulting from thermal fluctuations, as shown in Figure 8 (d). The switching probabilities from P to AP (P$_{01}$) and AP to P (P$_{10}$) are measured with various voltage durations, under an external field applied to compensate the stray field from the fixed layer. At the start of the pulse, a damped-oscillation behavior is observed, consistent with LLG dynamics [26,27]. The switching probability reaches 50 % when the pulse duration is approximately 3 ns, confirming that the switching is dominated by random thermal fluctuations.

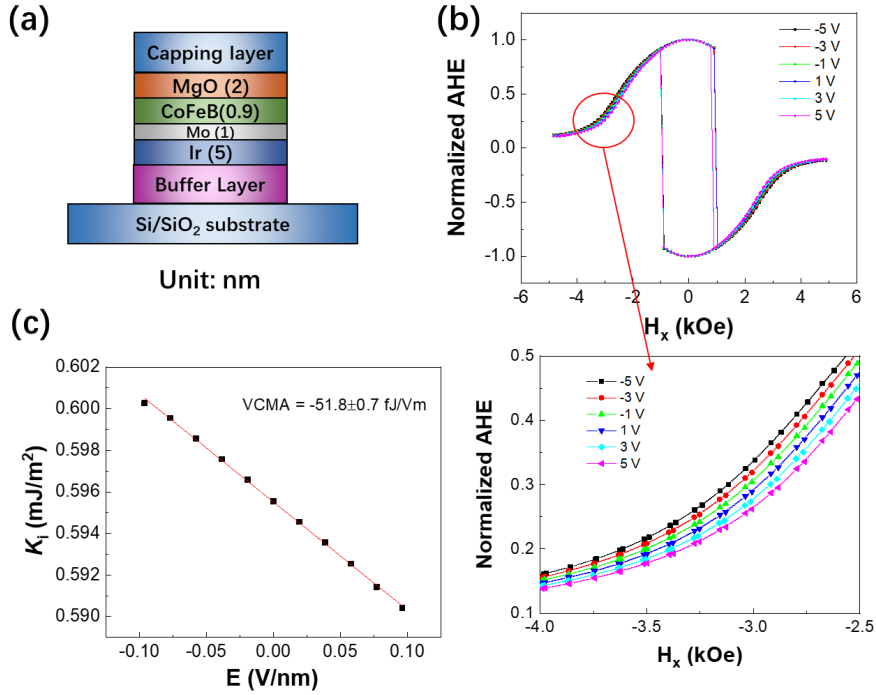

**Figure 7.** (a) Material stack for the VCMA Hall bar measurement (b) Normalized anomalous Hall resistance versus in-plane magnetic fields (hard-axis loops), under different gate voltages. We see a clear change in the anisotropy with different electric fields. (c) The extracted interfacial PMA values ($K_i$) as a function of applied electric field.

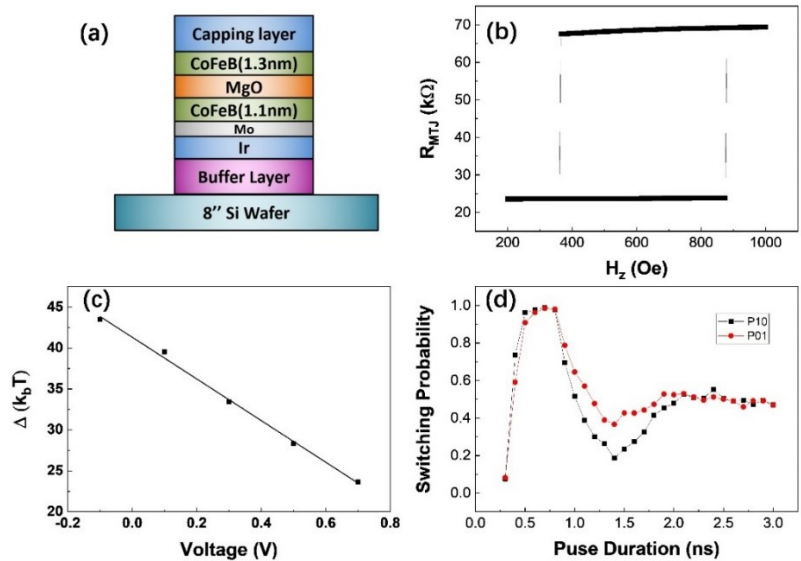

**Figure 8**. (a) Film stack of perpendicular MTJ. (b) Hysteresis loop of the MTJ. The shift in the magnetic R-H loop is due to the stray field of fixed layer. (c) Thermal stability dependence on applied voltage. (d) Switching probability of free layer with various pulse durations with the stray field compensated.

## 5. Conclusion

Thermodynamic computing provides a promising approach towards optimization by following nature's energy minimization process. In this work, we propose a magnetic thermodynamic core (MTC) processor based on VC-SOT MTJ devices. We derive the dynamics of each magnetic principle, the magnetic device, and the MTC; then demonstrate the computation capabilities of the MTC by using it to solve a 4-neuron Boltzmann Machine. Finally, we demonstrate strong SOT, VCMA, and thermal noise sensitivity in CoFeB-MgO devices. The demonstration paves the path towards the realization of efficient thermodynamic computing hardware.

It should be noted, however, that while this work demonstrates the use of our hardware for simulating a Boltzmann machine, the theoretical and algorithmic concepts of stochastic dynamics can reach far beyond that of this work. For example, it is well known that the noise level is crucial to the optimization process of stochastic processes, and correlates with its dynamic phase space [39,40]. A dynamically-adjusted stochastic system optimized with respect to its noise levels belongs to a finite-width phase that precedes ordinary chaotic behavior and deterministic computation [39]. Within the recently proposed supersymmetric theory of stochastic dynamics, this phase is a manifestation of the breakdown of topological supersymmetry [41,42], and is directly related to phenomena described in the literature as noise-induced chaos, self-organized criticality, and dynamical complexity [43]. It can be roughly interpreted as the noise-induced bridge between integrable and chaotic deterministic dynamics, and inherits the properties of both: on the one hand, the system has integrable flows with well-defined attractors that can be associated with candidate solutions, and on the other hand, the noise-induced attractor-to-attractor dynamics which are effectively chaotic/aperiodic in nature and avoid revisiting solutions/attractors, thus accelerating the search for the best solution. The ability to combine simulated and chaotic annealing leads to much faster convergence times to the global minimum and leads to efficient optimizers.

## Acknowledgements

The authors acknowledge the support from the National Science Foundation (NSF) ERC Center-TANMS, DARPA under FRANC, NSF ECE- 1935362, 1909416, and the Army Research Office Multidisciplinary University Research Initiative (MURI) under grant numbers W911NF16-1-0472. KLW also acknowledges the endowment of the Raytheon Chair.